\documentclass[preprint,aps,floats,showpacs,nofootinbib]{revtex4}
\usepackage{graphicx}
\usepackage{epsfig}
\usepackage{pstricks}
\usepackage{pst-coil}
\usepackage{amsmath}
\def\be{\begin{equation}}
\def\ee{\end{equation}}
\def\bea{\begin{eqnarray}}
\def\eea{\end{eqnarray}}
\begin{document}
\title{\bf Leptogenesis and low energy CP phases with two
heavy neutrinos}
\author{Kaushik Bhattacharya}
\email{kasuhik@prl.res.in}
\author{Narendra Sahu}
\email{narendra@prl.res.in}
\author{Utpal Sarkar}
\email{utpal@prl.res.in}
\author{Santosh K. Singh}
\email{santosh@prl.res.in}
\affiliation{Theoretical Physics Division, Physical Research Laboratory,
Navrangpura, Ahmedabad 380 009, India}
\begin{abstract}
An attractive explanation for non-zero neutrino masses and small matter 
antimatter asymmetry of the present Universe lies in ``leptogenesis". 
At present the {\it size} of the lepton asymmetry is precisely known, 
while the {\it sign} is not known yet. In this work we determine 
the sign of this asymmetry in the framework of two right handed 
neutrino models by relating the leptogenesis phase(s) with the 
low energy CP violating phases appearing in the leptonic mixing 
matrix. It is shown that the knowledge of low energy lepton number 
violating re-phasing invariants can indeed determine the sign of the 
present matter antimatter asymmetry of the Universe and hence 
indirectly probing the light physical neutrinos to be Majorana type.
\end{abstract}
\maketitle
\bigskip

\section{Introduction}
\label {intr}
Within the Standard Model (SM) the neutrinos are massless and hence 
there is no CP violation in the lepton (L) sector. The current 
evidence~\cite{solar-expt,atmos-expt,kamland} from the neutrino oscillation 
experiments, on the other hand, suggest that neutrinos are massive, 
however small, and they mix up. The goal of the present neutrino 
oscillation experiments is to determine the nine degrees of freedom 
in the low energy neutrino mass matrix. They are parametrized by three 
masses, three mixing angles and three CP violating phases out of which 
two are Majorana and one is Dirac. At present the neutrino oscillation 
experiments able to measure the two mass square differences, the solar 
and the atmospheric, and three mixing angles with varying degrees of 
precision, while there is no information about the phases. 

Assuming that the neutrinos are of Majorana type the small masses of
the physical left handed neutrinos can be explained by the elegant
seesaw mechanism~\cite{seesawgroup} which involves singlet right-handed  
neutrinos (type-I seesaw) or triplet Higgs (type-II
seesaw) or can be both (hybrid seesaw). In the present article
we limit ourselves to the case of type-I seesaw models. Although
we call them right-handed neutrinos, in the extensions of the
SM they are just singlet fermions that transform trivially under 
the SM gauge group. So, there is no
apparent reasons for the number of heavy singlet neutrinos to be 
same as the number of left-handed neutrinos. So, for the main 
part of our discussions we restrict ourselves to only two right-handed
neutrinos. These results will also be true when there are three
right-handed neutrinos, but the third right-handed neutrino
do not mix with the other two neutrinos. We start with 
three right-handed neutrinos and after some general comments work
mostly with two right-handed neutrinos.

While there is no information about the absolute mass scales of 
the physical neutrinos, the currently 
discovered tiny mass scales; the atmospheric neutrino mass 
($\Delta_{atm}=\sqrt{|m_3^2-m_2^2|}$) in the $\nu_{\mu}-\nu_{\tau}$ 
oscillation and the solar neutrino mass ($\Delta_{\odot}=\sqrt{m_2^2-m_1^2}$) 
in the $\nu_e-\nu_{\mu}$ oscillation, can be explained by adding at least 
two right handed neutrinos to the SM Lagrangian. However, with two 
right-handed neutrinos the seesaw mechanism predicts one of the physical light 
neutrino mass to be exactly zero which is permissible within the 
current knowledge of neutrino masses and mixings. 

The Majorana mass of the right-handed neutrino violates $L$-number and hence 
is a natural source of L-asymmetry in the early Universe~\cite{
fukugita.86,baryo_lepto_group}. A partial L-asymmetry is then converted 
to baryon (B) asymmetry through the non-perturbative sphaleron processes, 
unsuppressed above the electroweak phase transition. Currently the 
$B$-asymmetry has been measured precisely by the Wilkinson Microwave 
Anisotropy Probe (WMAP)\cite{spergel.03} and is given by
\begin{equation}
\left( \frac{(n_B-n_{\bar{B}})}{n_\gamma} \right)_0 \equiv
\left( \frac{n_B}{n_\gamma} \right)_0=\left(6.1^{+0.3}_{-0.2}\right)
\times 10^{-10}.
\label{asymmetry}
\end{equation}

It is legitimate to ask if there are any connecting links between 
leptogenesis and the CP violation in the low energy leptonic sector, 
in particular neutrino oscillation and neutrinoless double beta decay. 
In the context of three right-handed neutrino models several attempts have 
been taken in the literature to connect the CP violation in leptogenesis 
and neutrino oscillations~\cite{3RH-models}. It is found that there 
are almost no links between these two phenomena unless one considers 
special assumptions~\cite{scpv_models}. In fact it is shown 
that leptogenesis can be possible irrespective of the CP violation 
at low energy~\cite{rebelo_prd}. On the other hand, in the two right-handed 
neutrino models there is a ray of hope connecting leptogenesis with 
the CP violation in neutrino oscillation~\cite{2RH-models} and 
neutrinoless double beta decay processes. 

While the magnitude of CP violation is fairly known in the quark 
sector, it is completely shaded in the leptonic sector of the SM. 
Therefore, searching for CP violation in the leptonic sector is of 
great interest in the present days. It has been pointed out that the Dirac 
phase, being involved in the L-number conserving processes, can be 
measured in the long baseline neutrino oscillation experiments~\cite{
dirac_phase_group}, while the Majorana phase, being involved in the 
L-number violating processes, can be investigated in the neutrinoless 
double beta decay~\cite{majorana_phase_group} processes.

At present the magnitude of B-asymmetry is precisely known, while the 
sign of this asymmetry is not known yet. However, by knowing 
the CP violating phases in the leptonic mixing matrix one can determine 
the sign of the B-asymmetry. This is the study taken up in this work. We 
consider a minimal extension of the SM by including two singlet right-handed 
neutrinos which are sufficient to explain the present knowledge of 
neutrino masses and mixings. We adopt a general parameterization of the 
neutrino Dirac Yukawa coupling and give the possible links between the 
CP violation in leptogenesis and neutrino oscillation, CP violation 
in neutrinoless double beta decay and leptogenesis. It is shown that 
the knowledge of low energy CP violating re-phasing invariants can 
indeed determine the sign of the B-asymmetry since the size of this 
asymmetry is known precisely.   

Rest of the manuscript is arranged as follows. In section II we 
elucidate the canonical seesaw in the framework of three right handed 
neutrinos. We then display the possible links between leptogenesis and 
the low energy CP-violating phases appearing in the leptonic mixing matrix 
in certain special circumstances. It is found that there are almost no 
links between these two phenomena occurring at two different energy scales.
Therefore, in section III we give a parameterization of $m_D$ in the 
two right-handed neutrino models. In section IV we calculate the neutrino 
masses and mixings by using the parameterization of $m_D$ given in 
section III. In section V we estimate the CP violation in 
leptogenesis. In section VI we consider the re-phasing invariant 
formalism to study the possible links between the CP violating phases 
responsible for leptogenesis and the CP violation at low energy 
phenomena. First we calculate th CP violation in neutrino oscillation 
and then elucidate its link to leptogenesis. After that we calculate 
the CP violation in low energy lepton number violating process, i.e., the 
neutrinoless double beta decay, and then elucidate its link to 
leptogenesis. We conclude in section VII. 

\section{Canonical seesaw and parameter counting}
\label{sec2}
To account for the small neutrino masses we extend the SM by including 
right-handed neutrinos. The corresponding leptonic Lagrangian is given by
\begin{eqnarray}
\mathcal{L} &=& \overline{\ell_{Li}}i\gamma^\mu D_\mu\ell_{Li}+\overline{
\ell_{Ri}}i\gamma^\mu \partial_\mu\ell_{Ri}+\overline{N_{R\alpha}}i\gamma^\mu
\partial_\mu N_{R\alpha}\nonumber\\
&-&\left( {1\over 2}\overline{(N_{R\alpha})^c}(M_R)_{\alpha \beta}N_{R\beta}+
\overline{\ell_{Li}}
\phi (Y_e)_{ij}\ell_{Rj}+\overline{\ell_{Li}}\tilde{\phi} (Y_\nu)_{i\alpha} 
N_{R\alpha}+H.C.\right)\,,
\label{lagrangian}
\end{eqnarray}
where $\tilde{\phi}=i\tau^2\phi$ and $i$ runs from 1 to 3,
representing the left-handed fields. $\alpha$ represent the right
handed neutrino indices. $\ell_{Li}$ represents the ${\rm SU(2)}_L
\times {\rm U(1)}_Y$ doublets, $\ell_{Ri}$ and $N_{R\alpha}$ are
right-handed singlets of the theory.

After the electroweak symmetry breaking the canonical seesaw~\cite{
seesawgroup} gives the effective neutrino mass matrix
\begin{eqnarray}
m_\nu=-m_D M_R^{-1} m_D^T\,,
\label{seesaw}
\end{eqnarray}
where $m_D=Y_\nu v$ is the Dirac mass matrix of the neutrinos with 
$v$ is the vev of SM Higgs and that of $M_R$ is the mass matrix of 
right handed neutrinos. Without loss of generality we consider $M_R$ 
to be diagonal and in this basis $m_D$ contains rest of the physical 
parameters that appears in $m_\nu$. 

The diagonalization of $m_{\nu}$, through the lepton flavor mixing
matrix $U_{PMNS}$~\cite{pmns-matrix}, gives us three masses of the
physical neutrinos. Its eigenvalues are given by
\begin{eqnarray}
D_m\equiv {\rm diag.}(m_{1}, m_{2}, m_{3})=U_{PMNS}^\dagger
m_{\nu} U_{PMNS}^*\,,
\label{diag}
\end{eqnarray}
where the masses $m_i$ are real and positive. The standard PDG
parametrization~\cite{pdg} of the PMNS matrix reads:
\begin{eqnarray}
U_{PMNS}=
\left(
\begin{array}{ccccc}
c_{12}c_{13} & & s_{12}c_{13} & &s_{13}e^{-i\delta_{13}}\cr\\
-s_{12}c_{23}-c_{12}s_{23}s_{13}e^{i\delta_{13}} & & c_{12}c_{23}-
s_{12}s_{23}s_{13}e^{i\delta_{13}} & & s_{23}c_{13}\cr\\
s_{12}s_{23}-c_{12}c_{23}s_{13}e^{i\delta_{13}} & & -c_{12}s_{23}-s_{12}
c_{23}s_{13}e^{i\delta_{13}} & & c_{23}c_{13}
\end{array}
\right)\,.\,U_{ph}
\label{mns-matrix}
\end{eqnarray}
where $U_{ph}={\rm diag.}\,(1, e^{i\eta}, e^{i(\xi +\delta_{13})})$
and $c_{ij}$, $s_{ij}$ stands for $\cos \theta_{ij}$ and $\sin \theta_{ij}$
respectively. The two physical phases $\eta$ and $\xi$ associated with
the Majorana character of neutrinos are not relevant for neutrino
oscillations. Thus we see that there are three phases in the low energy 
effective theory responsible for CP violation. However, these phases may 
not give rise to CP violation at high energy regime, in particular, 
leptogenesis to our interest. In the following we study this in the 
framework of three and than two right-handed neutrino models. 

In general if $n$ and $n'$ are the number of generations of the 
left- and right-handed neutrinos that take part in the seesaw then 
the total number of parameters in the effective theory is estimated 
to be~\cite{broncano_plb.03}
\begin{eqnarray}
N_{\rm moduli}&=&n+n'+nn'\,,
\label{mod}\\
N_{\rm phase}&=&n(n'-1)\,.
\label{phs}
\end{eqnarray}
For $n=3$ and $n^\prime = 3$, $N_{\rm moduli} = 15$ and $N_{\rm phase}=6$,
which in the effective theory manifests as three masses of charged leptons, 
three masses of right-handed neutrinos and remaining 15 parameters including 
nine moduli and six phases in the Dirac mass matrix $m_D$ in a 
basis where the charged lepton mass matrix is real and diagonal. 

In the bi-unitary parameterization the mass matrix $m_D$ can be given 
as
\begin{equation}
m_D=U_L^\dagger m_D^{diag} U_R\,,
\label{para_step1}
\end{equation}
where $U_L$ and $U_R$ are $3\times 3$ unitary matrices. $U_L$ 
diagonalizes the left-handed sector while $U_R$ is the diagonalizing 
matrix of $m_D^\dagger m_D$. Any arbitrary $3 \times 3$ 
unitary matrix $U$ can be written as
\begin{eqnarray}
U=e^{i\varphi} P_1 {\widetilde U} P_2\,,
\label{udef}
\end{eqnarray}
where $\varphi$ is an overall phase and
\begin{eqnarray}
P_1 &=& {\rm diag.}(1, e^{-i\alpha_1}, e^{-i\alpha_2})\,,
\label{p1}\\
P_2 &=& {\rm diag.}(1, e^{-i\beta_1}, e^{-i\beta_2})\,,
\label{p2}
\end{eqnarray}
are phase matrices. ${\widetilde U}$ is a CKM like matrix parametrized
by three angles and one embedded phase. Now using Eq. (\ref{udef}) in  
Eq. (\ref{para_step1}) we get 
\be
m_D=e^{i(-\varphi_L+\varphi_R)}P_{2L}^\dagger {\widetilde U_L}^\dagger 
P_{1L}^\dagger m_D^{diag} P_{1R}{\widetilde U_R}P_{2R}.
\label{para_step2}
\ee
Without loss of generality three of the left phases can be absorbed in 
the redefinition of charged lepton fields. As a result the effective Dirac 
mass matrix turns out to be 
\be
m_D={\widetilde U_L}^\dagger P_3 m_D^{diag} {\widetilde U_R}P_{2R}\,, 
\label{para_step3}
\ee
where $P_3=P_{1L}^\dagger P_{1R}$ is an effective phase matrix. Thus 
in the models with three right-handed neutrinos $m_D$ contains 15 
parameters. 

In leptogenesis, the CP asymmetry comes in a form $m_D^\dagger m_D$, 
which contains $P_{2R}$ and ${\widetilde U_R}$, {\it i.e.}, 
\be
m_D^\dagger m_D=P_{2R}^\dagger {\widetilde U_R}^\dagger (m_D^{diag})^2 
{\widetilde U_R}P_{2R}\,,
\ee
and hence is independent of $P_{3}$ and ${\widetilde U_L}$. 
Although it would be good to know the exact relationship of the
phases in $P_{2R}$ and ${\widetilde U_R}$ with the phases appearing 
in the $U_{PMNS}$ matrix but that is not possible. So, we try with 
some special cases.

{\bf Case-I:} Let us first consider the case, when ${\widetilde U_R}$ 
is a diagonal matrix. This is the case when the right-handed neutrino 
Majorana mass 
matrix is diagonal to start with. The mass matrix can still contain 
Majorana phases. In that case, ${\widetilde U_R}$ and $m_D^{diag}$ 
will commute and hence $m_D^\dagger m_D$ will be real and there will 
not be any leptogenesis. This already tells us that the phases in 
leptogenesis crucially depends on the mixing of the right-handed 
physical neutrinos. Even in this case there will be CP violation 
at low energy as we shall see below. The light neutrino mass matrix 
is given by
\[ m_\nu = -{\widetilde U_L}^\dagger (P_3)^2({\widetilde U_R})^2 (P_{2R})^2 
(m_D^{diag})^2 M_R^{-1} {\widetilde U_L}^*\]
so that the PMNS matrix will become
\[ U_{PMNS} = {\widetilde U_L}^\dagger P_3P_{2R}\,.\]
Thus both the Dirac and Majorana phases at low energy are 
non-vanishing. 

{\bf Case-II:} We shall now consider another special case when there is no
leptogenesis. If the diagonal Dirac neutrino mass matrix is
proportional to a unit matrix, i.e.,
$m_D = m \cdot I$ ($I$ is the identity matrix),
again there is no leptogenesis, 
\[ m_D^\dagger m_D= m^2 \cdot I \,.\]
In this case the light neutrino mass matrix becomes
\[ m_\nu = -{\widetilde U_L}^\dagger P_3 {\widetilde U_R} P_{2R} 
m^2 M_R^{-1}P_{2R}{\widetilde U_R}^T P_3 {\widetilde U_L}^*\,,\]
so that the PMNS matrix can be read off to be
\[ U_{PMNS} = {\widetilde U_L}^\dagger P_3 {\widetilde U_R} P_{2R}\,. \]
Even in this case the Dirac and Majorana phases are present.

Thus in both these examples, even if CP violation is observed at low
energy neutrino experiments, this CP violation may not be related to 
leptogenesis. Since it is not possible to make any further progress 
with three heavy neutrinos, we shall now restrict ourselves to models 
with two heavy neutrinos. 

\section{Parameterization of $m_D$ in 2RH Neutrino models}
\label{sec3}
From now on we shall work with only two right-handed (2RH) neutrinos. 
This result will be applicable when there are only two heavy 
neutrinos or when there are three heavy neutrinos but one of them do 
not mix with others and heavier than the other right-handed neutrinos 
and hence its contribution to the light neutrinos is also negligible. 
In the present case where we have $n=3$ and $n'=2$, from Eq. (\ref{mod}) 
and (\ref{phs}), we get $N_{\rm moduli}=11$ and $N_{\rm phase}=3$. The 
14 parameters in the effective theory manifest them as three 
masses of charged leptons, two masses of right handed neutrinos and 
remaining nine parameters including six moduli and three phases appear 
in the Dirac mass matrix $m_D$.

There are various textures and their phenomenological implications of 
$m_D$ in the 2RH neutrino models that have been considered in the 
literature~\cite{2RH-textures}. In this article a general 
parametrization of the $3\times 2$ mass matrix of the Dirac neutrinos 
is considered. This is given by
\begin{eqnarray}
m_D=v Y_\nu=v U Y_{2RH}\,,
\label{trip}
\end{eqnarray}
where $U$ is an arbitrary Unitary matrix and the Yukawa coupling of the 
two RH neutrino model is given as 
\begin{eqnarray}
Y_{2RH}=
\left(
\begin{array}{ccc}
0 & & x\\
z && y e^{-i\theta}\\
0  & & 0
\end{array}\right)\,.
\label{tri1}
\end{eqnarray}
A derivation of Eq.~(\ref{tri1}) is given in the appendix \ref{appA}. 
However, we declare that the texture of $Y_{2RH}$ is not unique. 
By choosing appropriately the $U$ matrix one can place $x,y,z$ 
at different positions so as to get the different textures of 
$Y_{2RH}$ as shown in appendix \ref{appB}. Using (\ref{udef}) in 
Eq. (\ref{trip}) we get 
\begin{eqnarray}
m_D = v{\widetilde U} P_2 Y_{2RH}\,,
\label{emd}
\end{eqnarray}
where ${\widetilde U}$ contains four parameters including three 
moduli and one phase, $P_2$ contains two phases and $Y_{2RH}$ contains 
four parameters including three moduli and one phase which all together 
makes ten parameters in $m_D$. However, 
by multiplying the phase matrix $P_2$ with $Y_{2RH}$ one can see 
that one of the phases in the phase matrix $P_2$, i.e., $\beta_2$ becomes
redundant and can be dropped without loss of generality. 
As a result the total number of effective parameters is actually 
nine and hence consistent with our counting.

Substituting $m_D$, given by Eq. (\ref{emd}), in Eq. (\ref{seesaw}) 
we can calculate the effective neutrino mass matrix, $m_\nu$. The 
diagonalization of $m_{\nu}$, through the lepton flavor mixing 
matrix $U_{PMNS}$~\cite{pmns-matrix}, then gives us two non-zero 
masses of the physical neutrinos while setting one of the mass to be 
exactly zero as shown in the following section.
\section{Neutrino masses and mixings in 2RH neutrino models}
\label{sec4}
The unitary matrix ${\widetilde U}$, appearing in Eq.~(\ref{emd}), can 
be parameterized as~\footnote{This parameterization is usually used for 
determining the leptonic mixing matrix in the PDG parameterization. 
Here we have used it for parameterizing $m_D$.} 
\begin{eqnarray}
{\widetilde U} = R_{23}(\Theta_{23})R_{13}(\Theta_{13}, \delta'_{13})
R_{12}(\Theta_{12})\,.
\label{up}
\end{eqnarray}
It turns out that this parameterization is useful in determining the 
leptonic mixing matrix in 2RH neutrino models. Now from Eqs.~(\ref{seesaw}) 
and (\ref{emd}) we get the effective neutrino mass matrix to be
\begin{eqnarray}
m_\nu &=& -v^2 {\widetilde U} P_2 Y_{2RH} M_R^{-1} Y^T_{2RH} P_2 
{\widetilde U}^T\,\nonumber\\
&=& - v^2 {\widetilde U}P_2 X P_2 {\widetilde U}^T\,,
\label{ss1}
\end{eqnarray}
where 
\begin{eqnarray}
X= Y_{2RH} M_R^{-1} Y^T_{2RH}\,.
\label{xdef}
\end{eqnarray}
For simplicity of the calculation let us take $e^{-i\theta}$ common 
from 2nd row of $Y_{2RH}$ matrix given by Eq. (\ref{tri1}) and absorb 
it in $P_{2}$ by redefining $\beta_{1}$ as $(\beta_{1}+\theta)\rightarrow
\beta_{1}$. As a result opposite phase will reappear with $z$. Then the 
matrix $Y_{2RH}$ turns out to be
\begin{eqnarray}
Y_{2RH}=\left(\begin{array}{ccc}
0 &  & x\\
z e^{i\theta}&& y\\
0 &  & 0\end{array}\right)\,.
\label{tri2}
\end{eqnarray}

Using Eq. (\ref{tri2}) in the above Eq. (\ref{xdef}) we get   
\begin{eqnarray}
X=\left(\begin{array}{ccc}
{x^{2}\over M_{2}} & {xy\over M_{2}} & 0\\
{xy\over M_{2}} & {y^{2}\over M_2} +{z^{2}e^{2i\theta}\over M_{1}} & 0\\
0 & 0 & 0\end{array}\right)\,. 
\label{x}
\end{eqnarray}
In writing the above equation we have used a diagonal basis of 
the RH neutrinos where $M_R = {\rm diag.}(M_1, M_2)$. For simplicity, we 
absorb $M_{1}$ and $M_{2}$ in $x, y$ and $z$ as $\frac{x}{\sqrt{M_2}}
\rightarrow a, $ $\frac{y}{\sqrt{M_2}}\rightarrow b$ and $\frac{z}
{\sqrt{M_1}}\rightarrow c$. So $X$ can be rewritten as:
\begin{eqnarray}
X=\left(\begin{array}{ccc}
a^{2} & ab & 0\\
ab & b^{2}+c^{2}e^{2i\theta} & 0\\
0 & 0 & 0\end{array}\right)\,.
\label{nx} 
\end{eqnarray}
Looking to the effective neutrino mass matrix as given by Eq. (\ref{ss1}) 
we can guess that the diagonalizing matrix would be of the form 
\begin{eqnarray}
U_{PMNS}={\widetilde U} K\,,
\label{uk}
\end{eqnarray}
where $K$ is an unitary matrix. Using Eqs. (\ref{diag}) and (\ref{uk}) 
in Eq.~(\ref{ss1}) we see that
\begin{eqnarray}
D_m = - K^{\dagger}P_{2} X P_{2}K^*\,,
\label{dmk}
\end{eqnarray}
which implies that $K$ would diagonalize the matrix $P_{2} X P_{2}$. 
From the structure of $X$ it is clear that one of the light physical 
neutrinos must be massless. The matrix $K$ can be parameterized as
\begin{eqnarray}
K=P_2\,R_{12}(\omega,\phi)\,P\,,
\label{kpar}
\end{eqnarray}
where $P={\rm diag.}(e^{i\eta_{1}/2},e^{i\eta_{2}/2},1)$ and 
\begin{eqnarray}
R_{12}(\omega,\phi)=
\left(\begin{array}{ccc}
\cos \omega & e^{i\phi}\sin \omega & 0\\
-e^{-i\phi}\sin \omega & \cos \omega & 0\\
0 & 0 & 1\end{array}
\right)\,,
\label{rpar}
\end{eqnarray}
with
\begin{eqnarray}
\tan 2\omega &=&\left[\frac{2ab\left(a^4 + b^4+ c^4+ 2a^2b^2 +
2b^2c^2 \cos 2\theta+ 2c^2a^2 cos 2\theta\right)^{1/2}}
{\left( -a^{4}+b^{4}+c^{4}+2b^{2}c^{2}\cos 2\theta \right)}\right]\,,
\label{omeg}
\end{eqnarray}
and
\begin{eqnarray}
\tan \phi&=&\left[\frac{-c^{2}\sin 2\theta}{a^{2}+b^{2}+c^{2}\cos 2\theta}
\right]\,.
\label{ph}
\end{eqnarray}
Since $R_{12}(\omega,\phi)$ diagonalizes the matrix $X$ the resulting 
diagonal matrix will have complex eigenvalues in general. However, by 
choosing appropriately the phases of $P$ one can make the eigenvalues 
of $X$ real. Using Eqs. (\ref{omeg}) and (\ref{ph}) we get the 
eigenvalues $\{\lambda_1, \lambda_2, \lambda_3\}$ of $X$ to be 
\begin{equation}
\lambda_1=a^2-ab e^{i\phi}\tan \omega\,,~~\lambda_2=e^{-2i\phi}(a^2+
ab e^{i\phi}\cot\omega) ~~~{\rm and}~~~ \lambda_3=0
\label{lambda}
\end{equation} 
The absolute masses of the physical neutrinos are then given by $\{
m_1=v^2 |\lambda_1|, \, m_2=v^2 |\lambda_2|,\, m_3=0\}$.  
The MSW effect in the solar neutrino oscillation experiments indicates 
that $m_2 > m_1$. The corresponding mass scale, giving rise to the $\nu_e -
\nu_\mu$ oscillation, is given by
\begin{equation}
\Delta m^2_\odot\equiv m_2^2-m_1^2=v^4( |\lambda_2|^2-|\lambda_1|^2)\,.
\label{solar-mass}
\end{equation}
Using Eq. (\ref{lambda}) in the above equation we get the solar neutrino 
mass scale to be 
\begin{eqnarray}
\Delta m^2_\odot &=&v^4 \left\{ \left[ (a^2+b^2)^2+c^4+2b^2c^2\cos 2\theta 
\right]^2 - 4 a^4 c^4 \right\}^{1/2} \nonumber \\
&\simeq& 8\times 10^{-5} eV^2\,.
\label{sol-mass}
\end{eqnarray}
The atmospheric mass scale, on the other hand, is given by 
\begin{equation}
\Delta m^2_{atm} \equiv  |m_2^2 - m_3^2|=v^4( |\lambda_2|^2-|\lambda_3|^2)\,.
\end{equation}
Now using Eq. (\ref{lambda}) in the above equation we get the 
atmospheric mass scale to be
\begin{eqnarray}
\Delta m^2_{atm} &=& \frac{v^{4}}{2}\left((a^{2}+b^{2})^{2}
+c^{4}+2b^{2}c^{2}\cos2\theta \right. \nonumber \\ && +\left.
\left\{\left((a^{2}+b^{2})^{2}+
c^{4}+2b^{2}c^{2}\cos2\theta\right)^{2}-4a^{4}c^{4}\right\}^{1/2}
\right)\,,\nonumber\\
&\simeq & 2\times 10^{-3} eV^2\,.
\label{atmos-mass}
\end{eqnarray}
These equations may be inverted to obtain
\begin{eqnarray}
v^4 \left((a^{2}+b^{2})^{2}
+c^{4}+2b^{2}c^{2}\cos2\theta \right) &=& 2 \Delta m^2_{atm}
-\Delta m^2_\odot \nonumber \\
a^4 c^4 v^8 &=& \Delta m^2_{atm} (\Delta m^2_{atm}-\Delta m^2_\odot).
\end{eqnarray}

Now using Eqs. (\ref{p2}) and (\ref{rpar}) in Eq. (\ref{kpar}) 
we can rewrite the matrix $K$ as
\begin{eqnarray}
K &=& R_{12}(\omega,\phi+\beta_1)P'\nonumber\\
  &=& \left(\begin{array}{ccc}
\cos\omega & e^{i(\phi+\beta_{1})}\sin \omega & 0\\
-e^{-i(\phi+\beta_{1})}\sin \omega & \cos \omega & 0\\
0 & 0 & 1\end{array}\right)
\left(\begin{array}{ccc}
e^{i\eta_{1}/2} & 0 & 0\\
0 & e^{i(\eta_{2}/2-\beta_{1})} & 0\\
0 & 0 & e^{-i{\beta_2}}\end{array}
\right)\,.
\label{krlab}
\end{eqnarray}
Thus using Eqs. (\ref{krlab}) and (\ref{up}) in Eq. (\ref{uk}) the PMNS 
matrix $U_{PMNS}$ is given as
\begin{equation}
U_{PMNS}=R_{23}(\Theta_{23})R_{13}(\Theta_{13},\delta'_{13})R_{12}
(\Theta_{12}) R_{12}(w, \phi+\beta_1)P'\,,
\label{eff-pmns}
\end{equation}
where 
\begin{eqnarray}
&&R_{12}(\Theta_{12})R_{12}(\omega,\phi+\beta_{1})=\left(\begin{array}{ccc}
\cos\Theta_{12}'e^{i\rho_{1}} & \sin\Theta_{12}'e^{i\rho_{2}} & 0\\
-\sin \Theta_{12}'e^{-i\rho_{2}} & \cos\Theta_{12}'e^{-i\rho_{1}} & 0\\
0 & 0 & 1\end{array}\right)\nonumber\\
&=& \left(\begin{array}{lll}
e^{i\left(\frac{\rho_{1}+\rho_{2}}{2}\right)} & 0 & 0\\
0 & e^{-i\left(\frac{\rho_{1}+\rho_{2}}{2}\right)} & 0\\
0 & 0 & 1\end{array}
\right)\left(\begin{array}{lll}
\cos\Theta_{12}' & \sin \Theta_{12}' & 0\\
-\sin \Theta_{12}' & \cos \Theta_{12}' & 0\\
0 & 0 & 1\end{array}\right)
\left(\begin{array}{lll}
e^{i\left(\frac{\rho_{1}-\rho_{2}}{2}\right)} & 0 & 0\\
0 & e^{-i\left(\frac{\rho_{1}-\rho_{2}}{2}\right)} & 0\\
0 & 0 & 1\end{array}\right)\,.
\label{r12k}
\end{eqnarray}
In the above equation we have 
\begin{eqnarray}
\cos 2\Theta_{12}'&=&\cos 2\omega \cos 2\Theta_{12} - \cos(\phi+\beta_{1})
\sin 2\omega \sin 2\Theta_{12}\,,
\label{thetap}\\
\sin(\rho_{2}-\rho_{1})&=&\sin(\phi+\beta_{1})\tan \omega \left[
\cot 2\Theta_{12}'+\frac{\cos 2\Theta_{12}}{\sin 2\Theta_{12}' }\right]\,,
\label{2-1}\\
\sin(\rho_{1}+\rho_{2})&=&\frac{\sin 2\omega \sin(\phi+\beta_{1})}
{\sin 2\Theta_{12}'}\,.
\label{2+1}
\end{eqnarray}
For further simplification of the PMNS matrix (\ref{eff-pmns}) we now 
compute the matrix product $R_{12}(\Theta_{12})K=R_{12}(\Theta_{12})R_{12}
(\omega,\phi+\beta_{1})P'$ which is given as
\begin{eqnarray}
& &R_{12}(\Theta_{12})R_{12}(\omega,\phi+\beta_{1})P'= 
e^{i({\eta_1\over 2}-\rho_2)}\nonumber\\
& &\left(\begin{array}{ccc}
e^{i(\rho_{1}+\rho_{2})} & 0 & 0\\
0 & 1 & 0\\
0 & 0 & 1\end{array}\right)
\left(\begin{array}{ccc}
\cos\Theta_{12}' & \sin\Theta_{12}' & 0\\
-\sin\Theta_{12}' & \cos\Theta_{12}' & 0\\
0 & 0 & 1\end{array}\right)
\left(\begin{array}{ccc}
1 & 0 & 0\\
0 & e^{i(\rho_{2}-\rho_{1}+(\eta_{2}-\eta_{1})/2-\beta_{1})} & 0\\
0 & 0 & e^{-i(\beta_2-\rho_2+{\eta_1\over 2})}\end{array}
\right)\,.
\label{rkmult}
\end{eqnarray}
Now using Eq. (\ref{rkmult}) in Eq. (\ref{eff-pmns}) the $U_{PMNS}$ matrix 
can be rewritten as:
\begin{eqnarray}
U_{PMNS}&=&\tilde{U}K\nonumber\\
&=& R_{23}(\Theta_{23})R_{13}(\Theta_{13},\delta_{13})R_{12}
(\Theta_{12}')\, \nonumber\\
&&{\rm diag.}(1,e^{i(\rho_{2}-\rho_{1}+
(\eta_{2}-\eta_{1})/2-\beta_{1})}, e^{-i(\beta_2-\rho_2+{\eta_1\over 2})})
\nonumber\\
&=&V\,.\,V_{ph}\,,
\label{rdfu}
\end{eqnarray} 
where $V$ is the CKM like matrix and $V_{ph}$ is the Majorana phase matrix. 
The effective $CP$ violating phase in the $V$ matrix is given by
\begin{eqnarray}
\delta_{13}=\delta'_{13}+(\rho_{1}+\rho_{2})\,.
\label{dcp}
\end{eqnarray}
Note that in writing Eq.~(\ref{rdfu}) the overall phase
$e^{i({\eta_1\over 2}-\rho_2)}$ has been taken out.
Moreover, we absorb the
unphysical phase matrix ${\rm diag.}(1, e^{-(\rho_1+\rho_2)},
e^{-(\rho_1+\rho_2)})$ into the redefinition of charged lepton
fields. From Eqs.~(\ref{mns-matrix}), (\ref{up}) and (\ref{dcp}) we
see that, for the chosen parameterization of $Y_{2RH}$, two of the 
mixing angles $\Theta_{23}$ and $\Theta_{13}$ remains same as of the 
$(2-3)$ and $(1-3)$ mixing angles in PDG parameterization of the leptonic 
mixing matrix. Thus we can write $\Theta_{23}\equiv \theta_{23}$ and 
$\Theta_{13}\equiv\theta_{13}$. While $\Theta_{12}$ gets modified to 
$\Theta'_{12}$ and is given by Eq. (\ref{thetap}), the modified CP 
violating phase $\delta_{13}$ is given by Eq. (\ref{dcp}). At present 
the best fit value of $\Theta_{23}$ is given to be $45^\circ$, while 
the best fit value with $1\sigma$ error the value of $\Theta'_{12}$ 
is given to be $33.9^\circ \pm 1.6^\circ$~\cite{mohapatra_review.06}. 
The CHOOZ experiment gives a bound on $\Theta_{13}$. Currently the most 
conservative upper bound on $\Theta_{13}$ at the $3\sigma$ confidence 
level is given to be~\cite{fogli_group.05}
\begin{equation}
sin^2\Theta_{13}<0.048\,,
\end{equation}
which gives $\Theta_{13}<13^\circ$. 

\section{Leptogenesis in 2RH neutrino models}\label{sec5}
The Majorana mass of the RH neutrino violates L-number and hence is 
considered to be a natural source of L-asymmetry in the early 
Universe~\cite{fukugita.86} provided its decay violates CP symmetry, 
a necessary criteria of Sakharov~\cite{sakharov.67}. In a mass basis 
where the RH neutrinos are real and diagonal the Majorana neutrinos are 
defined as $N_i={1\over \sqrt{2}} (N_{Ri}\pm N_{Ri}^c)$. In this basis 
the CP asymmetry is given by
\begin{equation}
\epsilon_i=\frac{\Gamma_i-\bar{\Gamma_i}}{\Gamma_i+\bar{\Gamma_i}}\,,
\end{equation} 
where $\Gamma_i$ is the decay rate of $N_i$. If we assume a normal mass 
hierarchy ($M_1<<M_2$) in the RH neutrino sector then the final 
L-asymmetry is given by the decay of the lighter RH 
neutrino, $N_1$. The CP asymmetry parameter, arising from the decay of 
$N_1$, is then given by 
\begin{equation}
\epsilon_1={-3\over 16 \pi v^2}\left( {M_1\over M_2} \right) \frac{Im[ 
(m_D^\dagger m_D)_{12}]^2}{ (m_D^\dagger m_D)_{11} }\,.
\label{cpasym}
\end{equation}
Using Eqs. (\ref{emd}) and (\ref{tri1}) in the above Eq. (\ref{cpasym}) 
we get 
\begin{equation}
\epsilon_1={-3\over 16 \pi} \left( {M_1\over M_2}\right) y^2 \sin 2\theta\,.
\label{cpasym-1}
\end{equation}
From the above Eq. (\ref{cpasym-1}) it is clear that if $\theta=0$ then 
there is no CP violation in leptogenesis. Therefore, $\theta$ can be 
thought of the phase associated with $M_i$ in a basis where $M_i$'s 
are complex. Moreover, $\theta$ always hangs around $y$. So $y=0$ 
implies no leptogenesis. We will discuss more about it in sec.VI while 
we compare the CP violation in leptogenesis, neutrino oscillation and 
neutrinoless double beta decay processes.
\begin{figure}
\begin{center}
\epsfig{file=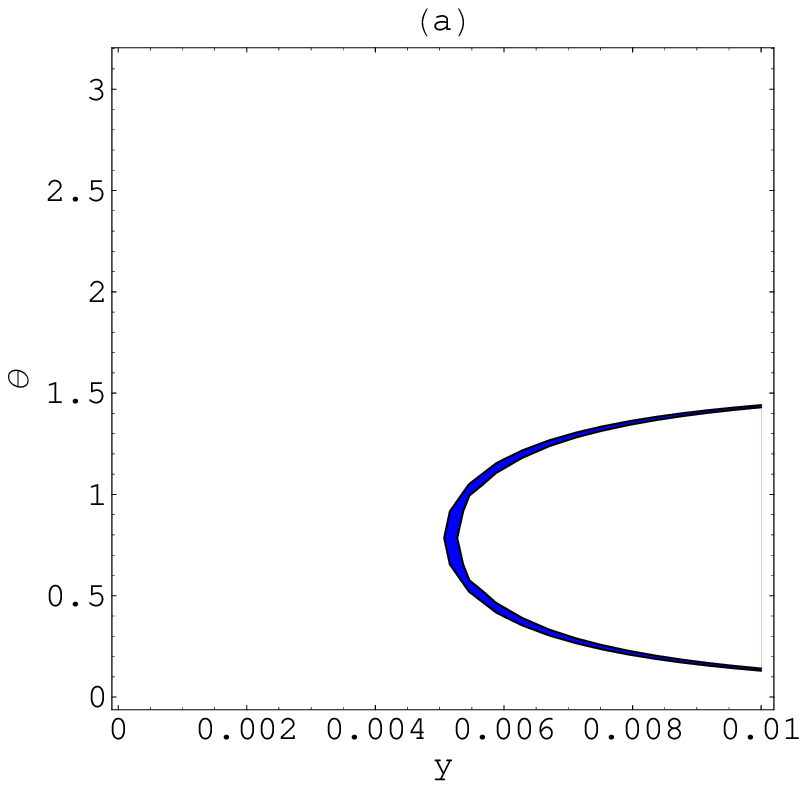, width=0.48\textwidth}
\epsfig{file=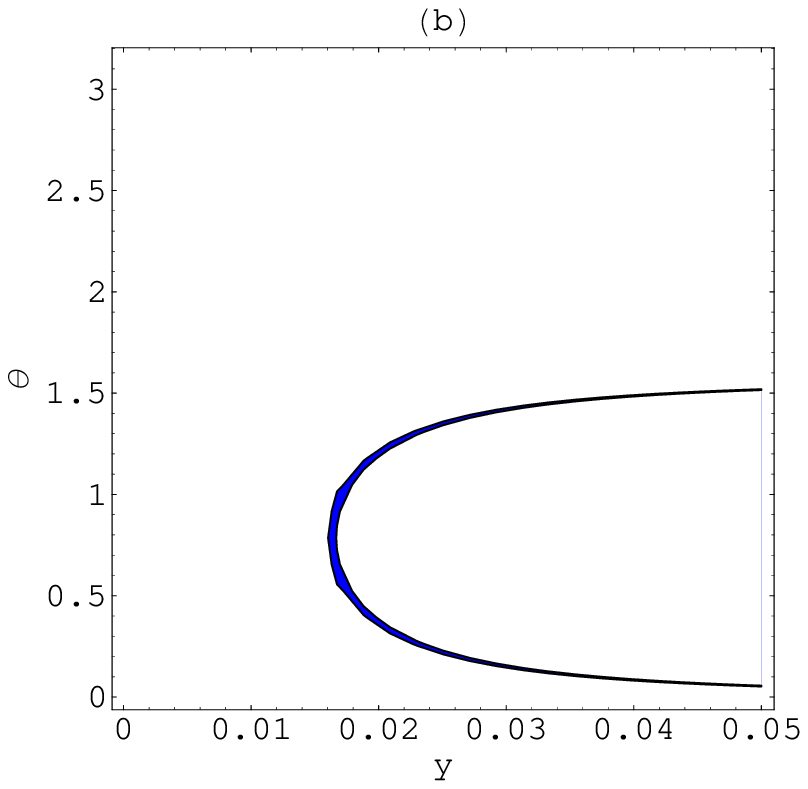, width=0.48\textwidth}
\end{center}
\caption{The allowed values of $y$ are shown against $\theta$ (in rad) for 
the observed matter anti-matter asymmetry, given by Eq. (\ref{asymmetry}), 
with (a)${M_1\over M_2}=0.1$ and (b)${M_1\over M_2}=0.01$.}
\label{ytheta}
\end{figure}

We now estimate the magnitude of $L$-asymmetry. A net L-asymmetry arises 
when $\Gamma_1$ fails to compete with the 
Hubble expansion parameter, $H=1.67g_*^{1/2}(T^2/M_{pl})$, where $g_*$ 
is the number of relativistic degrees of freedom at the epoch of 
temperature $T$. In a comoving volume the $L$-asymmetry is defined as 
\begin{equation}
Y_L=\epsilon_1 Y_{N_1} d\,,
\label{YL-def}
\end{equation}  
where $d$ is the dilution factor arises due to the competitions between 
$\Gamma_1$ and $H$ at $T\simeq M_1$. Now using Eq. (\ref{cpasym-1}) in 
the above Eq. (\ref{YL-def}) we get 
\begin{equation}
Y_L=-5.97 \times 10^{-5}{M_1\over M_2} \left( \frac{Y_{N_1}d}{10^{-3}}\right)
y^2 \sin 2\theta\,.
\label{cal-YL}
\end{equation}
A part of the L-asymmetry is then transferred to the B-asymmetry via the 
sphaleron processes which are unsuppressed above the electroweak 
phase transition. Taking into account the particle content in the $SM$ 
the B- and L-asymmetries are related as 
\begin{equation}
B=\frac{p}{p-1}L\simeq -0.55 L\,,
\label{BL-rel}
\end{equation} 
where $p=28/79$ appropriate for the particle content in the $SM$. As a 
result we get the net B-asymmetry per comoving volume to be 
\begin{equation}
Y_B\simeq 3.28 \times 10^{-5}{M_1\over M_2} \left( 
\frac{Y_{N_1}d}{10^{-3}}\right) y^2 \sin 2\theta\,.
\label{com-asy}
\end{equation}
The observed B-asymmetry, on the other hand, is given by
\begin{equation}
\left( \frac{n_B}{n_\gamma} \right) = 7 Y_B = 
2.3\times 10^{-4} {M_1\over M_2} \left( \frac{Y_{N_1}d}{10^{-3}}\right) 
y^2 \sin 2\theta \,.
\label{final-asy}
\end{equation}
Comparing the above Eq. (\ref{final-asy}) with the observed matter 
antimatter asymmetry, given by Eq.(\ref{asymmetry}), we get 
\begin{equation}
y^2 \sin 2\theta=(2.57~~-~~2.78)\times 10^{-6}{M_2\over M_1}
\left( \frac{10^{-3}}{Y_{N_1}d} \right)\,.
\label{theory-expt}
\end{equation}
We have shown the allowed values of $y$ in fig. (\ref{ytheta}), 
using $(Y_{N1}d)=10^{-3}$, for hierarchical RH neutrinos in the $y-\theta$ 
plane. It is shown in fig. \ref{ytheta}(a) that for $(M_1/M_2)=0.1$ the 
minimum allowed value of $y$ is $5\times 10^{-3}$. However, this value is 
lifted up to $1.7\times 10^{-2}$ for $(M_1/M_2)=0.01$ as shown in 
fig. \ref{ytheta}(b).

\section{CP violation in re-phasing invariant formalism}
It is convenient to study CP violation in a re-phasing invariant
formalism. In particular, for the CP violation in the leptonic sector
the latter makes it very interesting. The CP violation in any lepton number
conserving processes comes out to be of the form~\cite{jarlskog}
\begin{equation}
J_{abij} = {\rm Im} [ V_{ai} V_{bj} V^\ast_{aj} V^\ast_{bi} ],
\end{equation}
where $V$ is the CKM like matrix in the lepton sector. On the other 
hand, CP violation in any lepton number violating processes will be 
of the form~\cite{Nieves:1987pp}
\begin{equation}\label{x1}
t_{aij} = {\rm Im} [ V_{ai} V^\ast_{aj} (V_{ph})^\ast_{ii} (V_{ph})_{jj}]\,.
\end{equation}
Now one can have as many independent re-phasing invariant measures $t$ as 
many independent Majorana CP phases. For three generations there are two 
independent $t$'s (denoted as $J_1$ and $J_2$) and one J (denoted as 
$J_{CP}$). For example, in the neutrinoless double beta decay 
the following re-phasing invariant will appear
\begin{equation}\label{x2}
    T = {\rm Im} [ V_{ai} V_{aj} V^\ast_{bi} V^\ast_{bj} ] \sim
    t_{aij} t^\ast_{bij}.
\end{equation}
It has been shown that the re-phasing invariant CP violating quantity
$J_{CP}$ only appears in the neutrino oscillations and that of $J_1$, 
$J_2$ appears in the neutrinoless double beta decay processes which may 
be observed in the next generation experiments.

\subsection{CP-violation in leptogenesis and neutrino oscillation}
It has been pointed out that the Dirac phase $\delta_{13}$ can be 
measured in the long baseline neutrino oscillation 
experiments~\cite{dirac_phase_group}. In that case the CP violation 
arises from the difference of transition probability 
$\Delta P= P_{\nu_e\rightarrow \nu_\mu}-P_{\bar{\nu}_e\rightarrow 
\bar{\nu}_\mu}$. It can be shown that the transition probability 
$\Delta P$ is proportional to the leptonic Jarlskog invariant
\begin{equation}
J_{CP}=Im [V_{e1}V^*_{e2}V^*_{\mu 1}V_{\mu 2}]\,.
\label{jcp}
\end{equation} 
\begin{figure}
\begin{center}
\epsfig{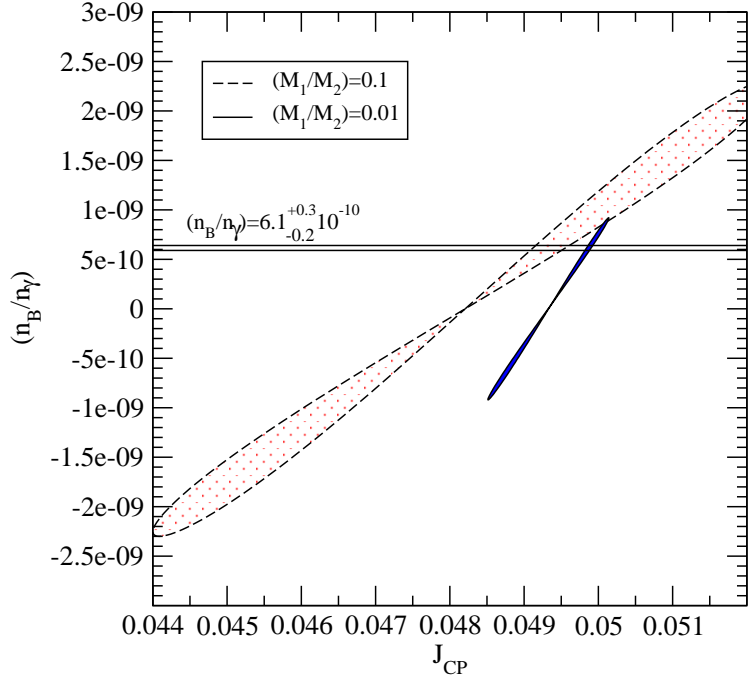}
\end{center}
\caption{The overlapping region in the $n_B/n_\gamma - J_{CP}$ plane is 
shown as $\theta$ (in rad) varies from $0$ to $\pi$ with
$\Theta_{23}=\pi/4$, $\Theta_{13}=13^\circ$, $\delta'_{13}=\beta_1=\pi/2$ 
and $z=x=0.01$. The dashed line is obtained for $\Theta_{12}=33.5^\circ$,
$y=0.01$ and ${M_1\over M_2}=0.1$, while the solid line is obtained for
$\Theta_{12}=33.8^\circ$, $y=0.02$ and ${M_1\over M_2}=0.01$.}
\label{jcp-fig}
\end{figure}
\begin{figure}
\begin{center}
\epsfig{file=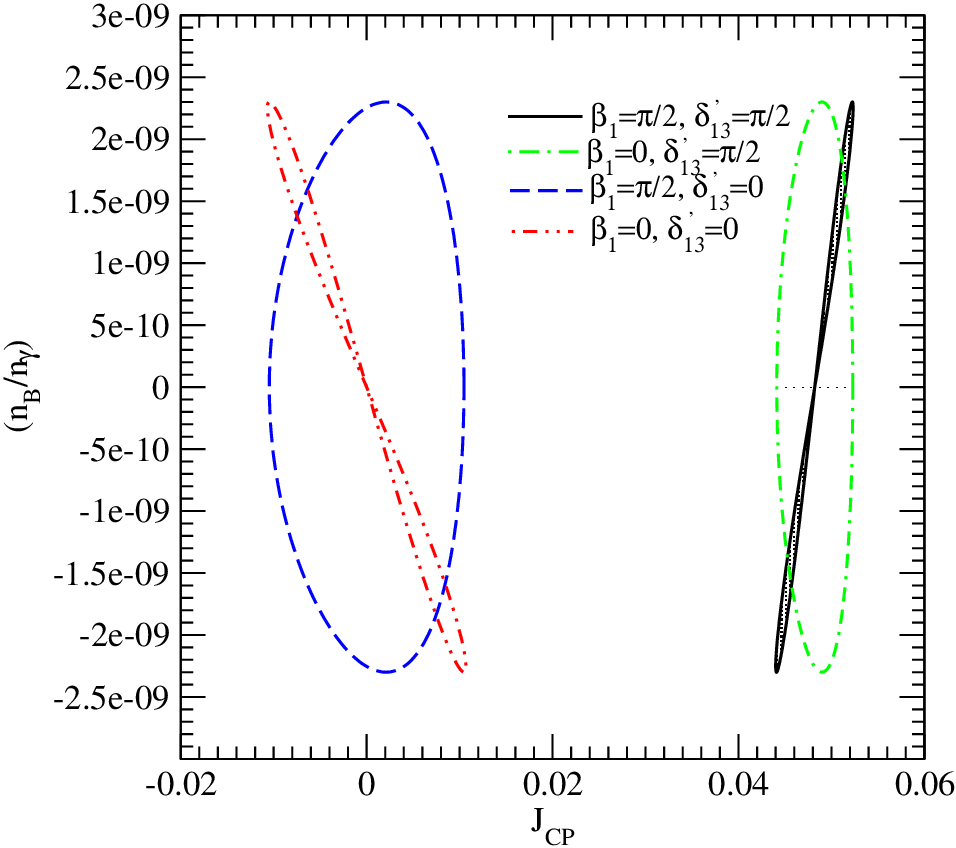, width=0.6\textwidth}
\end{center}
\caption{The variation of $n_B/n_\gamma$ is shown against $J_{CP}$ 
for different values of $\beta_1$ and $\delta'_{13}$ as $\theta$ 
(in rad) varies from $0$ to $\pi$. We have chosen $\Theta_{23}=\pi/4$, 
$\Theta_{13}=13^\circ$, $\Theta_{12}=33.5^\circ$, $x=y=z=0.01$ 
and ${M_1\over M_2}=0.1$.}
\label{jcpvalues-fig}
\end{figure}

Using Eq. (\ref{rdfu}) the re-phasing invariant $J_{CP}$ can be rewritten as 
\begin{equation}
J_{CP}=\frac{1}{8}\sin 2\Theta'_{12}\sin 2\Theta_{23}\sin 2\Theta_{13}
\cos \Theta_{13} \sin (\delta'_{13}+\rho_1+\rho_2)\,.
\label{jcp-1}
\end{equation}
Now using Eqs. (\ref{omeg}), (\ref{ph}), (\ref{thetap}) and (\ref{2+1}) in 
the above Eq. (\ref{jcp-1}) we get 
\begin{eqnarray}
J_{CP} &=& \frac{1}{8}\frac{\sin 2\Theta_{23}\sin 2\Theta_{13}\cos \Theta_{13}}
{\sqrt{\left[(a^2+b^2)^2+c^4+2b^2c^2\cos 2\theta\right]^2-4a^4c^4}}\nonumber\\
&\times& \left[ 2ab \cos\delta'_{13}\{ -c^2 \sin 2\theta 
\cos\beta_1+(a^2+b^2+c^2\cos2\theta)\sin\beta_1\}\right.\nonumber\\
&+&\left. 2ab\cos 2\Theta_{12}\sin\delta'_{13}\{ (a^2+b^2+c^2\cos 
2\theta)\cos\beta_1+c^2\sin 2\theta \sin\beta_1\}\right.\nonumber\\
&+& \left.\sin \delta'_{13}\sin 2\Theta_{12}(-a^4+b^4+c^4+2b^2 c^2 
\cos 2\theta) \right]\,.
\label{jcp-2}
\end{eqnarray}
From the above Eq. (\ref{jcp-2}) it is obvious that $J_{CP}=0$ only if 
both $\sin \delta'_{13}=0$ and $b=0$, while only $b=0$ (equivalently $y=0$) 
implies the condition for ``no leptogenesis". This indicates that there 
is no one-to-one correspondence between the CP violation in neutrino 
oscillation and the CP violation in leptogenesis, even in the 2RH 
neutrino models. However, it is interesting to see the common regions 
in the plane of $(n_B/n_\gamma)$ versus $J_{CP}$. This is shown 
in fig. (\ref{jcp-fig}) by taking a typical set of parameters. The 
main aim is to illustrate the maximal contrast between the positive 
and negative values of $n_B/n_\gamma$ for a given set of values of 
$J_{CP}$. This helps us in determining the sign of the asymmetry 
by knowing the size of the asymmetry. From the fig. (\ref{jcp-fig}) 
it is obvious that for the given set of parameters the positive sign 
of the asymmetry allows the values of $J_{CP}$ in the range $0.049 - 
0.0495$ for $(M_1/M_2)=0.1$. However, this range is significantly 
reduced for $(M_1/M_2)=0.01$. On the other hand, the negative sign of 
the asymmetry allows the values of $J_{CP}$ in the range $0.0465 - 0.047$ 
for $(M_1/M_2)=0.1$ which is further reduced for $(M_1/M_2)=0.01$. In 
this figure the value of $\Theta_{12}$ is used from 
fig. (\ref{theta12-theta}) where we have shown the allowed values of 
$\Theta_{12}$ as $\theta$ varies from $0$ to $\pi$. Note that the above 
results are true for a non-zero $\Theta_{13}$. Consequently the allowed 
range of values of $J_{CP}$ may vary depending on the values of 
$\Theta_{13}$. Thus we anticipate that in the 2RH neutrino models a 
knowledge of $J_{CP}$ can predict the sign of matter antimatter asymmetry 
of the Universe. We should note that the predictive power of the model 
depends on the CP violating phases $\beta_1$ and $\delta'_{13}$. This can 
be visible from fig. (\ref{jcpvalues-fig}) where we have shown the 
variation of $n_B/n_\gamma$ with $J_{CP}$ for different values of 
$\beta_1$ and $\delta'_{13}$. In particular, for the choice 
($\beta_1=\pi/2$, $\delta'_{13}=0$) and ($\beta_1=0$, 
$\delta'_{13}=\pi/2$), the contrast between the positive and negative 
values of $n_B/n_\gamma$ is almost zero for a given set of values of 
$J_{CP}$. On the other hand, for the choice ($\beta_1=\pi/2$, 
$\delta'_{13}=\pi/2$) and ($\beta_1=0$, $\delta'_{13}=0$), the contrast 
between the positive and negative values of $n_B/n_\gamma$ is maximal 
and can be chosen for the present purpose. 
\subsection{CP violation in leptogenesis and neutrinoless double-beta decay} 
The observation of the neutrinoless double beta decay would provide 
direct evidence for the violation of total $L$-number in the low 
energy effective theory and hence probing the left-handed physical 
neutrinos to be Majorana type. Note that the $L$-number violation at high 
energy scale is a necessary criteria for leptogenesis. In the canonical 
seesaw models this is conspired by assuming that the RH neutrinos are 
Majorana in nature. However, this assumption doesn't ensure that the 
left-handed physical neutrinos are Majorana type. Assuming that the 
physical neutrinos are of Majorana type we investigate the connecting 
links between the two $L$-number violating phenomena occurring 
at two different energy scales. 

In the low energy effective theory with three generations of left-handed 
fermions, apart from the $J_{CP}$, one can write two more re-phasing 
invariants $J_1$ and $J_2$ which designates lepton number violation and 
CP violation~\cite{Nieves:1987pp}. However, in the models with two RH 
neutrinos one of the eigen values of the physical light neutrino mass matrix 
is exactly zero. Therefore, the corresponding phase in the Majorana phase 
matrix can always be chosen so as to set one of the lepton number 
violating CP violating re-phasing invariant to zero. In the present 
case $m_3=0$ and hence the corresponding phase is arbitrary. This is 
ensured through $\beta_2$ which is redundant and pointed out in 
Eq. (\ref{emd}). Therefore, from Eq. (\ref{rdfu}) we can write 
the only $L$-number violating CP violating re-phasing invariant as: 
\begin{eqnarray}
J&=& Im \left[ V_{e1}V_{e2}^*(V_{ph})_{11}^* (V_{ph})_{22}\right]\nonumber\\ 
&=& -\frac{1}{2}\sin 2\Theta'_{12} \cos^2\Theta_{13} \sin(\rho_2-\rho_1+
{(\eta_2-\eta_1)\over 2}-\beta_1)\,. 
\label{J_lep}
\end{eqnarray}
\begin{figure}
\begin{center}
\epsfig{file=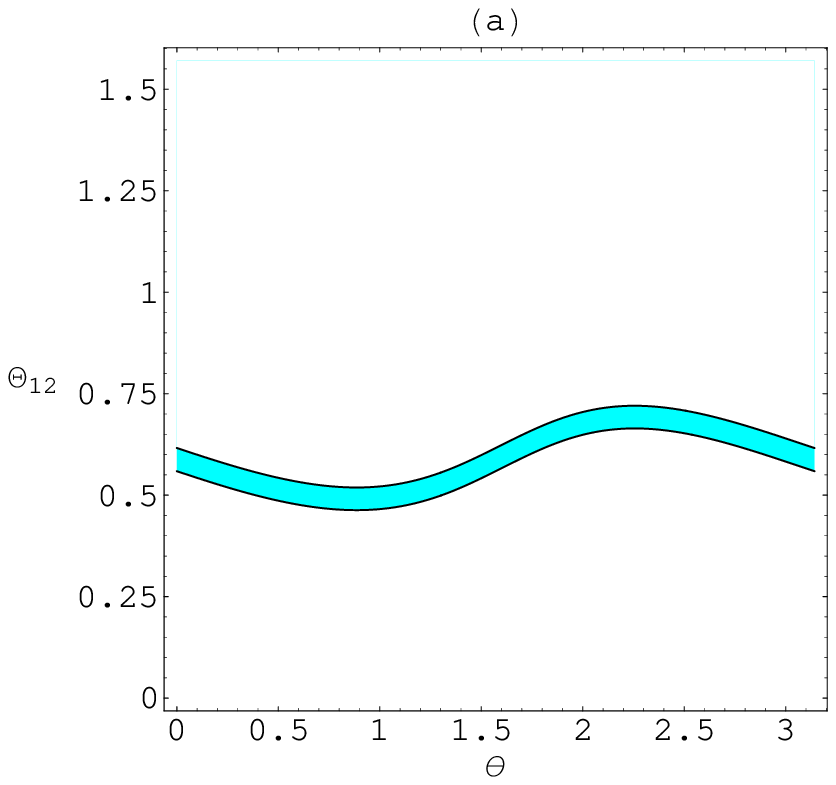, width=0.45\textwidth}
\epsfig{file=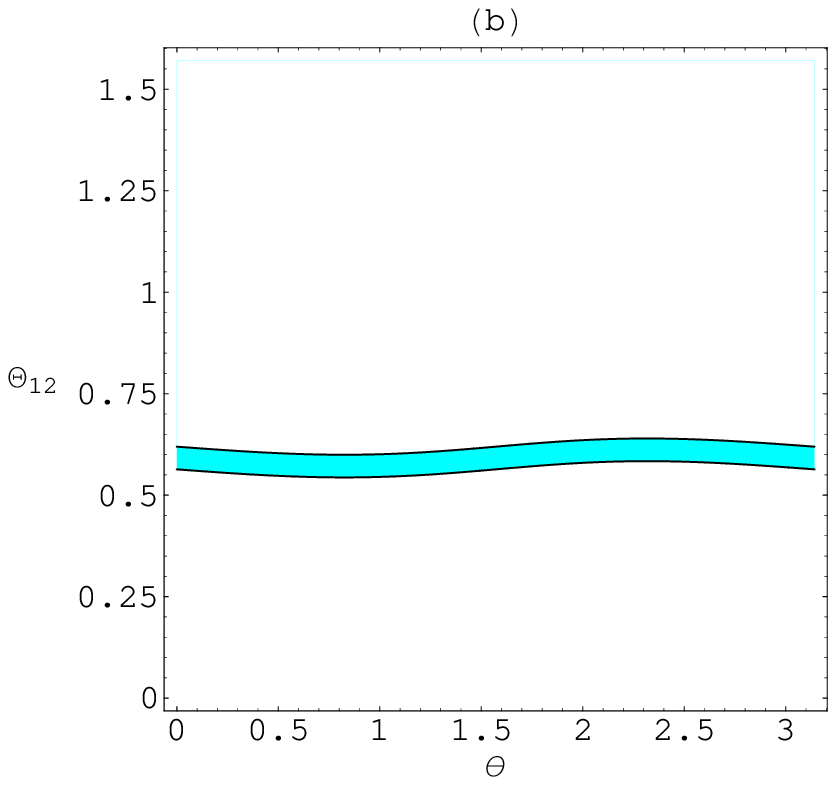, width=0.45\textwidth}
\caption{The allowed range of $\Theta_{12}$ (in rad) in 
Eq. (\ref{theta12_prime}) is shown as $\theta$ (in rad) varies from 
$0$ to $\pi$ for $\Theta'_{12}=(33.9\pm 1.6)^\circ$, $\beta_1=\pi/2$, 
$x=z=0.01$ (a) $y=0.01$ and $(M_1/M_2)=0.1$, and (b) $y=0.02$ 
and $(M_1/M_2)=0.01$.}
\label{theta12-theta}
\end{center}
\end{figure}
Using Eq. (\ref{rkmult}) the above Eq. (\ref{J_lep}) can be rewritten as 
\begin{eqnarray}
J &=&-\frac{\cos^2 \Theta_{13}}{2}\frac{1}
{\left[ (a^2+b^2)^2+c^4+2b^2c^2 \cos 2\theta+2c^2a^2 \cos 2\theta 
\right]}\nonumber\\
&\times& \left[ \sin 2\Theta_{12} \cos \theta \{-c^2 \sin 2\theta \cos 
\beta_1+(a^2+b^2+c^2 \cos 2\theta) \sin \beta_1 \}\right.\nonumber\\
&&\left. \times \sqrt{(a^2+b^2)^2+c^4+2c^2a^2+2b^2c^2 \cos 2\theta}\right.
\nonumber\\
&+& \left. \sin 2\Theta_{12} \sin \theta \{c^2 \sin 2\theta \sin \beta_1+ 
(a^2+b^2+c^2 \cos 2\theta) \cos \beta_1 \}\right.\nonumber\\
&& \left. \times \frac{(-a^4+b^4+c^4+2 b^2c^2 \cos 2\theta)}{\sqrt{(a^2+b^2)^2
+c^4+2 c^2a^2+2b^2c^2 \cos 2\theta}}\right.\nonumber\\
&+&\left. \cos 2\Theta_{12}\sin \theta \frac{2ab\{(a^2+b^2)^2+c^4+2b^2c^2 
\cos 2\theta +2c^2a^2 \cos 2\theta \}}{\sqrt{(a^2+b^2)^2+c^4+2c^2a^2+
2b^2c^2 \cos 2\theta}} \right]\,.
\label{J-2-final}
\end{eqnarray}
In the above Eq. (\ref{J-2-final}) the allowed values of $\Theta_{12}$ 
is obtained from 
\begin{eqnarray}
\cos \Theta'_{12} &=& \left[\frac{1}{2}\left[1+\frac{\left(-a^{4}+b^{4}+
c^{4}+2b^{2}c^{2}\cos2\theta\right)\cos 2\Theta_{12}}{\sqrt{((a^2+b^2)^2+c^4+
2b^2c^2 \cos 2\theta)^2-4a^4c^4}}\right.\right.\nonumber\\
&&-\left.\left.\sin2\Theta_{12} \frac{2ab\{ c^{2}\sin2\theta\sin\beta_{1}
+(a^{2}+b^{2}+c^{2}\cos2\theta)\cos\beta_{1}\} }{\sqrt{((a^2+b^2)^2+c^4+
2b^2c^2 \cos 2\theta)^2-4a^4c^4}}\right] \right]^{1/2}\,,
\label{theta12_prime}
\end{eqnarray}
by fixing $\Theta'_{12}=(33.9\pm 1.6)^\circ$. This is shown in 
fig. (\ref{theta12-theta}).
\begin{figure}
\begin{center}
\epsfig{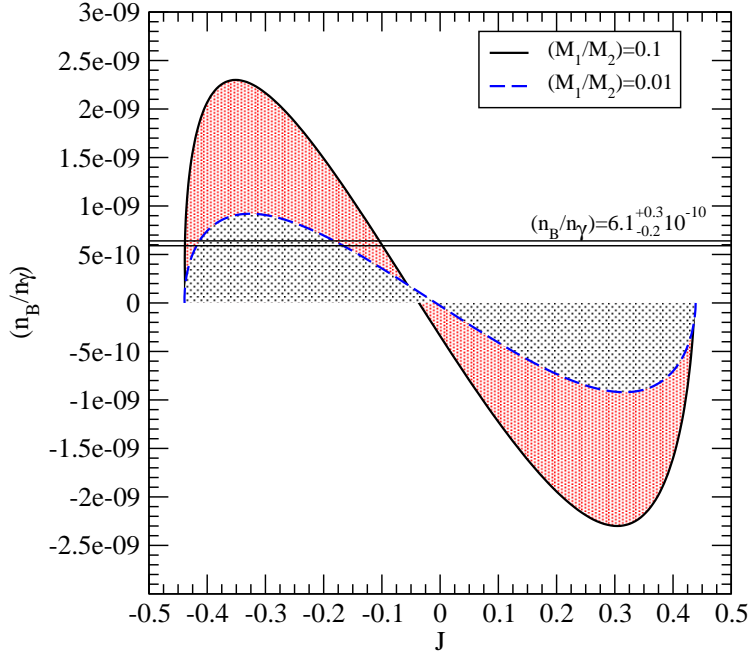}
\caption{The overlapping region in the $\frac{n_B}{n_\gamma} - J$ plane 
is shown as $\theta$ (in rad) varies from $0$ to $\pi$ with 
$\Theta_{13}=13^\circ$, $\beta_1=\pi/2$ and $x=z=0.01$. The solid line 
is obtained for $y=0.01$ and $\Theta_{12}=33.5^\circ$, while the dashed 
line is obtained with $y=0.02$ and $\Theta_{12}=33.8^\circ$.}
\label{j2-asy}
\end{center}
\end{figure}
\begin{figure}
\begin{center}
\epsfig{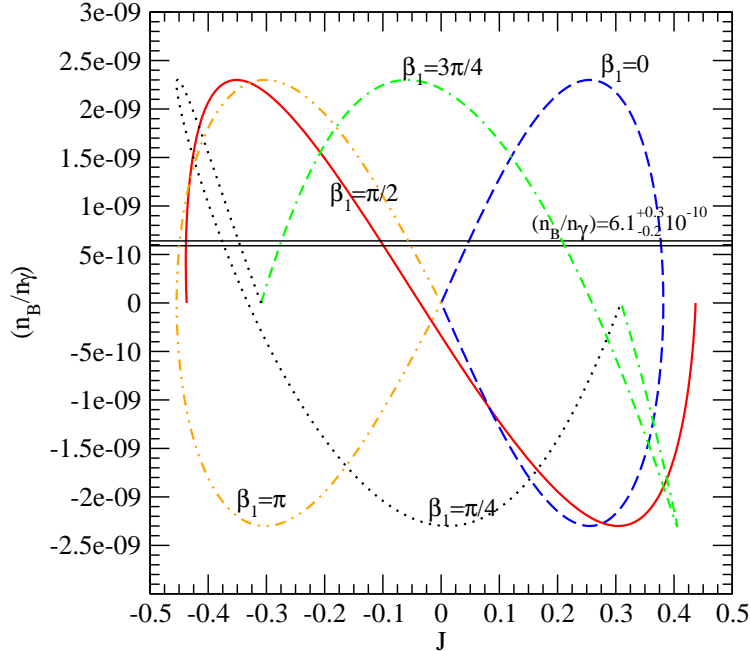}
\caption{The variation of $\frac{n_B}{n_\gamma}$ is shown against 
$J$ for different values of $\beta_1$ as $\theta$ (in rad) varies 
from $0$ to $\pi$. We have chosen $\Theta_{13}=13^\circ$, $\Theta_{12}=
33.5^\circ$, $x=y=z=0.01$ and $\frac{M_1}{M_2}=0.1$.}
\label{jvalues-asy}
\end{center}
\end{figure}

From Eq. (\ref{J-2-final}) one can see that $J \neq 0$ as $\theta
\rightarrow 0$ which is the condition for ``no leptogenesis". Thus 
we see that there is no one-to-one correspondence between the two 
$L$-number violating processes occurring at two different energy scales. 
However, it is always interesting to see the overlapping regions in the 
plane of $\frac{n_B}{n_\gamma}$ versus $J$ as $\theta$ varies from 
$0$ to $\pi$. This is shown in fig. (\ref{j2-asy}) for a typical 
set of parameters. From fig. (\ref{j2-asy}) one can see that 
for positive sign of the $B$-asymmetry the values of $J$ lie in between 
$-0.45$ to $-0.1$ for $(M_1/M_2)=0.1$. This range is further reduced to 
$(-0.4 - -0.15)$ for $(M_1/M_2)=0.01$. On the other hand, for the negative
sign of the $B$-asymmetry the values of $J$ lie in the range 
$(0.05 - 0.45)$ for $(M_1/M_2)=0.1$ and in the range $(0.15 - 0.4)$ 
for $(M_1/M_2)=0.01$. Thus we see that within the allowed range of 
parameters the contrast between the positive and negative values of 
$\frac{n_B}{n_\gamma}$ is maximum for a given set of values of $J$. 
Therefore, we expect a knowledge of $J$ can precisely determine the 
sign of $B$-asymmetry since the value of $n_B/n_\gamma$ is known. 
Finally we note that, unlike $J_{CP}$, $J$ remains non-vanishing even 
if $\Theta_{13}=0$~\footnote{In three generations there are two 
of them. See for example the paper by Y. Liu and U. Sarkar in 
ref.~\cite{Nieves:1987pp}}. Now the remaining question to be 
addressed is how $n_B/n_\gamma$ varies with respect to $J$ for 
different values of $\beta_1$. This is shown in fig. (\ref{jvalues-asy}) 
for a given set of parameters. One can see that for $\beta_1=0$ and 
$\beta_1=\pi$ both positive and negative values of $n_B/n_\gamma$ 
correspond to the same set of values of $J$ which is unwelcome for 
determination of sign of the asymmetry. On the other hand for 
$\beta_1\neq 0,\pi$ one can have maximal contrast between the positive 
and negative values of $n_B/n_\gamma$ for the given set of values of 
$J$ and hence can be chosen for the present purpose.     
\section{Conclusions}
We have studied the connecting links between the CP violating phase(s)
giving rise to leptogenesis, occurring at a high energy scale,
and the CP violating phases appearing in the low energy phenomena, i.e.,
neutrino oscillation and neutrinoless double beta decay processes. This 
is studied in the framework of two right-handed neutrino models. The low 
energy leptonic CP violation is studied in a re-phasing invariant formalism. 
It is shown that there are only two re-phasing invariants; (1) The lepton 
number conserving CP violating re-phasing invariant $J_{CP}$ which can be 
determined in the future long-baseline neutrino oscillation experiments, 
(2) The lepton number violating CP violating re-phasing invariant $J$ which 
can be determined in the neutrinoless double beta decay experiments. It is 
found that there is no one-to-one correspondence between these two CP 
violating phenomena, occurring at two different energy scales, even though 
the number of parameters involving in the seesaw is exactly same as the 
number of low energy observable parameters. However, in a suitable parameter 
space we have shown that the overlapping regions in the plane of 
$n_B/n_\gamma$ versus $J_{CP}$ and $n_B/n_\gamma$ versus $J$ can indeed 
determine the {\it sign} of the matter antimatter asymmetry of the 
present Universe assuming that the {\it size} of the asymmetry is 
precisely known. 

\begin{appendix}
\section{Parameterization of $Y_{2RH}$}
\label{appA}
To parameterize the neutrino Dirac Yukawa coupling in two right-handed 
neutrino models we follow the same procedure adopted in 
Ref.~\cite{morozumietal}. 
Let ${\bf u_1}$, ${\bf u_2}$, ${\bf u_3}$ be three orthonormal 3 
dimensional vectors. Using these basis vectors we can write the most 
general unitary matrix $U$ as:
\begin{eqnarray}
U = ({\bf u_1}\,\,{\bf u_2}\,\,{\bf u_3})\,.
\label{udef1}
\end{eqnarray} 
Let us consider an arbitrary $3 \times 2$ matrix $Y$ which in terms of 
the 3-dimensional vectors ${\bf y_1}$ and ${\bf y_2}$ can be written as:
\begin{eqnarray}
Y = ({\bf y_1}\,\,{\bf y_2})\,.
\label{ydef}
\end{eqnarray} 
Without loss of generality we choose ${\bf u_2}=\frac{{\bf y_1}}
{|{\bf y_1}|}$. As a result we get
\begin{eqnarray}
U^\dagger Y =
\left( 
\begin{array}{cccc}
0& & \alpha_{12}\\
|y_1| & & \alpha_{22}\\
0& & \alpha_{32} 
\end{array}
\right)\,,
\label{udagy}
\end{eqnarray}
where $\alpha_{ij}={\bf u_i}^\dagger \cdot {\bf y_j}$. 

Let $V$ be another unitary matrix which we choose to be of the form:
\begin{eqnarray}
V=
\left(
\begin{array}{ccccc}
\frac{\alpha_{12}}{\sqrt{|\alpha_{12}|^2+|\alpha_{32}|^2}} & & 0 & & 
\beta_{13}\\
0 & & 1 & & 0 \\
\frac{\alpha_{32}}{\sqrt{|\alpha_{12}|^2+|\alpha_{32}|^2}} & & 0 & & 
\beta_{33}
\end{array}
\right)\,,
\label{vmat}
\end{eqnarray}
where $\beta_{ij}$ must follow $\alpha_{12} \beta^*_{13}+\alpha_{32}
\beta^*_{33}=0$ and $|\beta_{13}|^2+|\beta_{33}|^2=1$.
Consequently we have
\begin{eqnarray}
V^\dagger U^\dagger Y =
\left(
\begin{array}{cccc}
0& & \sqrt{|\alpha_{12}|^2 + |\alpha_{32}|^2}\\ 
|y_1| & & \alpha_{22}\\
0  & & 0
\end{array}
\right)\,,
\end{eqnarray}
where we set $V_{32}=0$ by imposing the unitarity condition of $V$. This 
implies that we can always write any arbitrary $3\times 2$ matrix 
\begin{eqnarray}
Y = W Y_{2RH}\,,
\end{eqnarray}
where $W=VU$ is an unitary matrix and the texture of $Y_{2RH}$, the Yukawa 
coupling in the two right handed neutrino mass models, is given 
as
\begin{eqnarray}
Y_{2RH}=
\left(
\begin{array}{cccc}
0 & & x\\
z& & y e^{-i\theta}\\
0  & & 0
\end{array}\right)\,,
\end{eqnarray}
where $x$, $y$, $z$ and $\theta$ are real numbers. Note that by 
appropriately choosing the $U$ and $V$ matrices one can construct 
the $Y_{2RH}$ matrix in twelve possible ways. 
\section{Possible textures of $Y_{2RH}$ and neutrino mixings}
\label{appB}
In this appendix we specify the various possible textures of
$Y_{2RH}$. One of the particular texture of $Y_{2RH}$ has been used in
section \ref{sec2} for our work. In the table-I we write all the
possible textures of $Y_{2RH}$.
\begin{table}
\caption{Possible textures of $Y_{2RH}$}
\begin{center}
\begin{tabular}{|c|c|c|}
\hline 
\multicolumn{3}{|c|}{Zeros in the first row}\\[2mm]\hline
I & 
 $\begin{pmatrix} 0 & 0\cr
z & y e^{-i\theta}\cr
0 & x \end{pmatrix}$ \, 
$\begin{pmatrix} 0 & 0\cr
0 & x\cr
z & ye^{-i\theta} \end{pmatrix}$ & 
$\begin{pmatrix} 0 & 0\cr
x & 0 \cr
y e^{-i\theta} & z\end{pmatrix}$ \, 
$\begin{pmatrix} 0 & 0\cr
y e^{-i\theta} & z\cr
x & 0 \end{pmatrix}$\\[2mm]\hline
\multicolumn{3}{|c|}{Zeros in the middle row}\\[2mm]\hline
II & $\begin{pmatrix} z & y ^{-i\theta}\cr
0 & 0\cr
0 & x \end{pmatrix}$ \, 
$\begin{pmatrix} 0 & x\cr
0 & 0\cr
z & y e^{-i\theta}\end{pmatrix}$ & 
$\begin{pmatrix}y ^{-i\theta} & z\cr
0 & 0\cr
x & 0 \end{pmatrix}$
$\begin{pmatrix} x & 0\cr
0 & 0\cr
y e^{-i\theta} & z \end{pmatrix}$\\[2mm]\hline
\multicolumn{3}{|c|}{Zeros in the bottom row}\\[2mm]\hline
III & 
$\begin{pmatrix} z & y e^{-i\theta}\cr
0 & x\cr
0 & 0 \end{pmatrix} $\,
$\begin{pmatrix} 0 & x\cr
z & y e^{-i\theta}\cr
0 & 0\end{pmatrix} $ &

$\begin{pmatrix} y e^{-i\theta} & z\cr
x & 0\cr
0 & 0 \end{pmatrix}$ \,
$\begin{pmatrix} x & 0\cr
y e^{-i\theta} & z\cr
0 & 0 \end{pmatrix}$\\[2mm]\hline

\end{tabular}
\end{center}
\end{table}
Each possible $Y_{2RH}$ in table-I will lead to various forms
of $X$, apparent from Eq.~(\ref{xdef}). Accordingly the neutrino 
masses and mixing angles will be modified through the $m_D$ parameters.
\end{appendix}
\section*{Acknowledgment} It is our pleasure to thank Prof. Anjan 
Joshipura for helpful discussions. 

\end{document}